\documentstyle[psfig,sprocl]{article}

\def\ETAL{{\em et al.}}

\def\be{\begin{equation}}
\def\ee{\end{equation}}
\def\bea{\begin{eqnarray}}
\def\eea{\end{eqnarray}}

\begin{document}

\title{STRANGE CONTENT OF THE NUCLEON (NuTeV)}

\author{ 
 T.~ADAMS$^{4,*}$, {A.~ALTON$^{4}$}, {S.~AVVAKUMOV$^{7}$},
 L.~de~BARBARO$^{5}$, P.~de~BARBARO$^{7}$, R.~H.~BERNSTEIN$^{3}$,
 A.~BODEK$^{7}$, T.~BOLTON$^{4}$, J.~BRAU$^{6}$, D.~BUCHHOLZ$^{5}$,
 H.~BUDD$^{7}$, L.~BUGEL$^{3}$, J.~CONRAD$^{2}$, R.~B.~DRUCKER$^{6}$,
 R.~FREY$^{6}$, {J.~FORMAGGIO$^{2}$},{J.~GOLDMAN$^{4}$}, 
 {M.~GONCHAROV$^{4}$},
 D.~A.~HARRIS$^{7}$, R.~A.~JOHNSON$^{1}$, S.~KOUTSOLIOTAS$^{2}$,
 {J.~H.~KIM$^{2}$}, M.~J.~LAMM$^{3}$, W.~MARSH$^{3}$, 
 {D.~MASON$^{6}$}, {C.~MCNULTY$^{2}$}, K.~S.~MCFARLAND$^{7}$, 
 D.~NAPLES$^{4}$, P.~NIENABER$^{3}$, {A.~ROMOSAN$^{2}$}, W.~K.~SAKUMOTO$^{7}$,
 H.~SCHELLMAN$^{5}$, M.~H.~SHAEVITZ$^{2}$, P.~SPENTZOURIS$^{2}$ , 
 E.~G.~STERN$^{2}$, {B.~TAMMINGA$^{2}$}, {M.~VAKILI$^{1}$},
 {A.~VAITAITIS$^{2}$}, {V.~WU$^{1}$}, {U.~K.~YANG$^{7}$}, J.~YU$^{3}$ and 
 {G.~P.~ZELLER$^{5}$}}

\address{
 $^{*}$Presented by T. ADAMS \\
 $^{1}$University of Cincinnati, Cincinnati, OH, USA \\            
 $^{2}$Columbia University, New York, NY, USA \\                   
 $^{3}$Fermi National Accelerator Laboratory, Batavia, IL, USA \\  
 $^{4}$Kansas State University, Manhattan, KS, USA \\              
 $^{5}$Northwestern University, Evanston, IL, USA \\               
 $^{6}$University of Oregon, Eugene, OR, USA \\                    
 $^{7}$University of Rochester, Rochester, NY, USA                 
}


\maketitle\abstracts{The NuTeV experiment uses neutrino
deep-inelastic scattering from separate neutrino and anti-neutrino
beams to study the structure of the nucleon.  Charged-current
production of charm is sensitive to the strange content of the
nucleon while neutral-current charm production probes the charm
content.  Preliminary analyses of both topics are presented along
with discussion of possible momentum asymmetry in the strange
sea.}

\begin{figure}
 \hspace{1cm} \psfig{figure=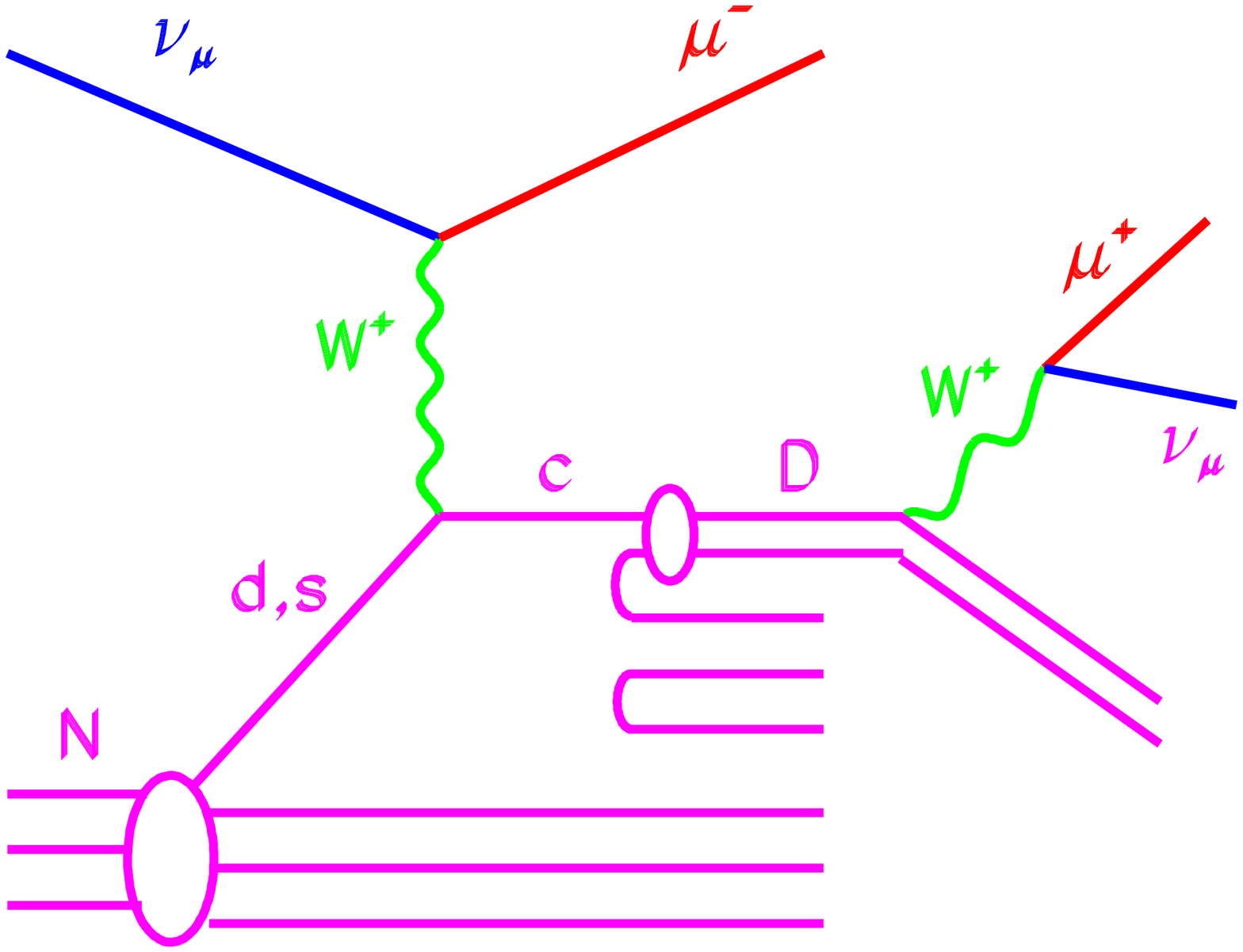,width=5cm} \hfill
 \psfig{figure=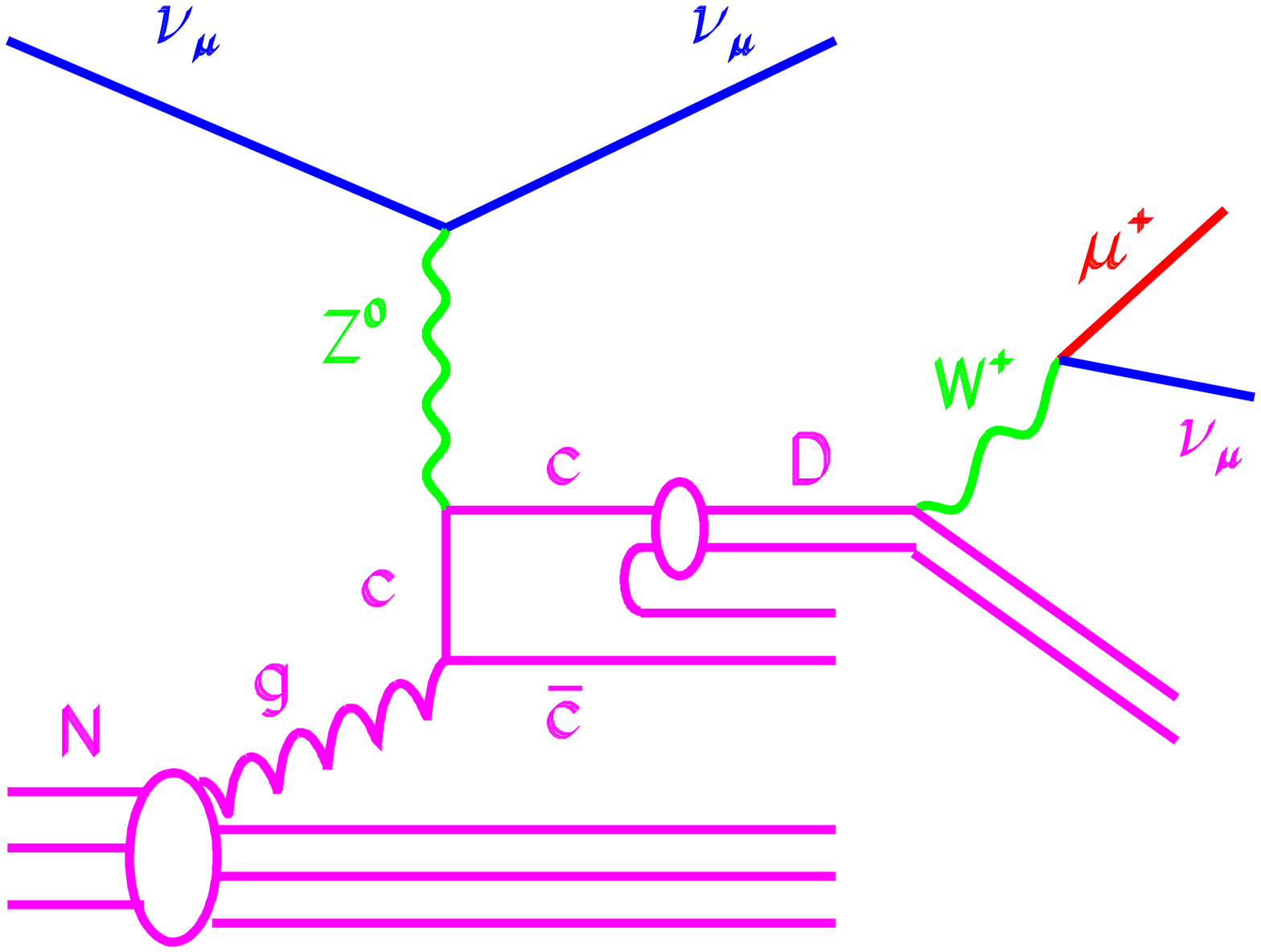,width=5cm} \hspace{1cm}

 \hspace*{2cm} (a) \hfill (b) \hspace{3cm}
 \caption{Feynman diagrams for (a) charged-current and (b) neutral-current charm
production via neutrino DIS.
 \label{fig:tadams:feyn_diagrams}}
\end{figure}

Charm production is a significant fraction of the total neutrino
deep-inelastic scattering (DIS) cross-section.  The semi-muonic
decay of charm mesons creates unique final states with which to study 
exclusive
charm production.  For charged-current reactions, an opposite-signed
two muon (dimuon) final state is possible.  Neutral-current production
can create an event with a single muon with charge opposite that which is
expected from a charged-current interaction (wrong-signed
muon).  Feynman diagrams for both reactions are shown in 
Fig.~\ref{fig:tadams:feyn_diagrams}.  

The NuTeV experiment studies $\nu$ and $\bar{\nu}$ DIS
using the Sign-Selected Quadru\-pole Train (SSQT) beam and the
Fermilab Lab E detector.\cite{bib:tadams:labe}  The SSQT allows 
separate running
of neutrino and anti-neutrino beams.  The (anti-)neutrino beam is incident
upon the Lab E target/calorimeter.  The target/calorimeter has 42 
segments each consisting
of two liquid-scintillator counters and one drift chamber interspersed
in 20 cm of iron.  The calorimeter provides energy and position measurement for
hadronic/electromagnetic showers and a position measurement for
deeply-penetrating muon tracks.
Immediately downstream of the calorimeter is a toroid spectrometer which 
provides 
momentum and charge measurement for the muons entering it.  The NuTeV 
analyses presented here
use the full data sample from the Fermilab 1996-97 fixed 
target run.  

\section{Charged-Current Charm Production \label{sec:tadams:cccharm}}

Neutrino charged-current DIS charm production results from scattering off
$d$ or $s$ quarks in the nucleon.  The Cabibbo suppression of 
scattering off $d$ quarks greatly enhances the contribution of
the $s$ quarks.
This allows the strange content of the nucleon to be probed.

The dimuon data set consists of events passing fiducial and kinematic
selections and containing a hadronic shower ($>$ 10 GeV) with at least 
two muons.  One muon must be toroid analyzed with more than 9 GeV in
energy while the other muon must have more than 5 GeV in energy.
The two sources of such events are charged-current charm production and a
charged-current event with a $\pi/K$ decaying within the
shower.  Both sources are included within the dimuon Monte Carlo
simulation.


Data and Monte Carlo are binned in three variables: 
\begin{eqnarray}
x_{vis} & = & 
\frac{E_{vis} E_{\mu 1} \theta^2_{\mu 1}}{2 m_p (E_{\mu 2} + E_{had})} \\
E_{vis} & = & E_{\mu 1} + E_{\mu 2} + E_{had} \\
z_{vis} & = & \frac{E_{\mu 2}}{E_{\mu 2} + E_{had}}
\end{eqnarray}
where the subscript $vis$ ($visible$) refers to measured variables, $\mu 1$ is
the primary muon (from the primary vertex) and $\mu 2$ is the secondary 
muon (from the decay vertex).  The 
Monte Carlo is weighted by the leading-order (LO) cross-section and
normalized to agree with the two-muon data.

Four parameters are used to describe charged-current charm
production.
$\kappa$ determines the size of the strange sea relative to the
non-strange sea ($\kappa \sim \frac{2\bar{s}}{\bar{u}+\bar{d}}$).   The shape
of the strange sea is described by $(1-x)^\beta$.  $m_c$ is the
mass of the charm quark and $\epsilon$ determines the 
Collins-Spiller fragmentation.~\cite{bib:tadams:collins}
These parameters are varied to find the best agreement between data
and Monte Carlo.

\begin{figure}
 \centerline{\psfig{figure=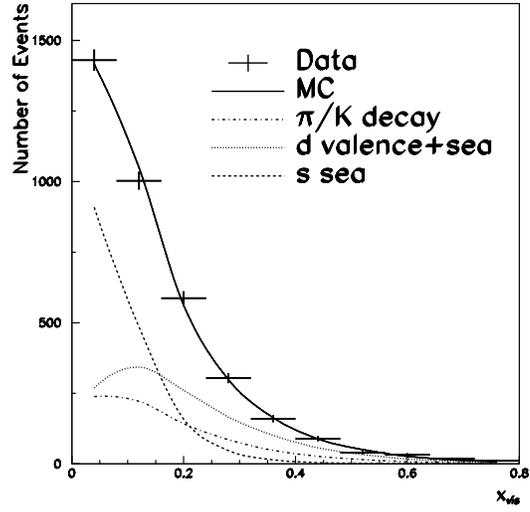,width=7.cm}}
 \centerline{(a)}
 \centerline{\psfig{figure=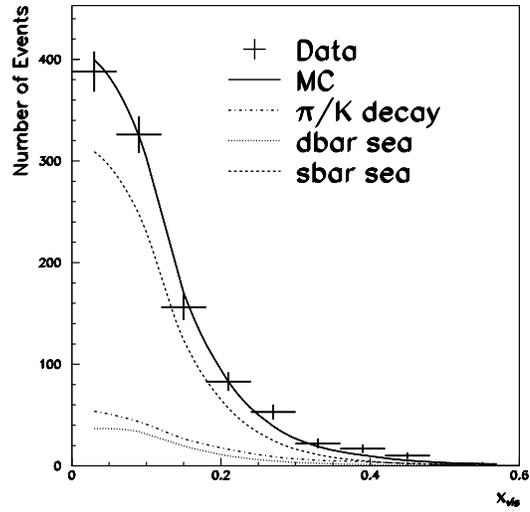,width=7.cm}}
 \centerline{(b)}

 \caption{Comparison of the NuTeV data with results of the fit for
the variable $x_{vis}$ in both $\nu$ (a) and $\bar{\nu}$ (b) mode.
 \label{fig:tadams:dimuxdist}}
\end{figure}
\begin{figure}
 \psfig{figure=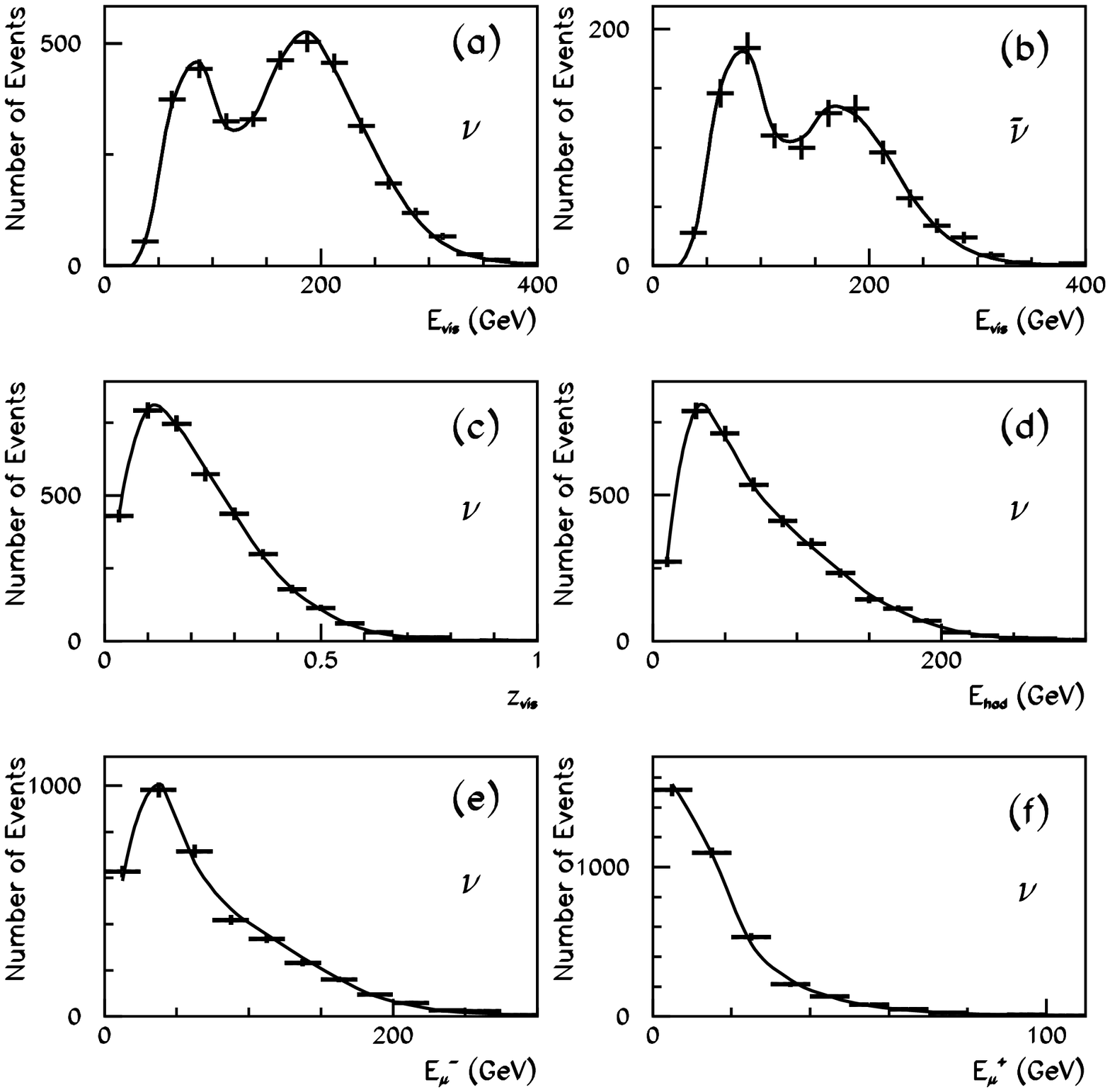,width=12.cm} \hfill
 \caption{Comparison of the NuTeV data with results of the fit
for several variables:  
$E_{vis}$ ($\nu$ mode) (a); $E_{vis}$ ($\bar{\nu}$ mode) (b); 
$z_{vis}$ ($\nu$ mode) (c); $E_{HAD}$ ($\nu$ mode) (d); 
$E_{\mu^-}$ ($\nu$ mode) (e); $E_{\mu^+}$ ($\nu$ mode) (f); 
 \label{fig:tadams:dimucomp}}
\end{figure}

Figures~\ref{fig:tadams:dimuxdist} and ~\ref{fig:tadams:dimucomp} 
show comparisons of data and Monte Carlo after fitting.  The 
distributions of $x_{vis}$ are shown in 
Fig.~\ref{fig:tadams:dimuxdist} with a breakdown of the components
of the simulation.  The dominance of production off the strange
sea can be clearly seen.
There is excellent agreement for various other distributions 
which have been compared, some of which are shown in 
Fig~\ref{fig:tadams:dimucomp}.
 
The preliminary results of the fit are:
\bea
 \kappa   & = & 0.42 \pm 0.07 \pm 0.06 \nonumber \\
 \beta    & = & 8.5 \pm 0.56 \pm 0.39 \hspace{0.5cm} (at~Q^2=16~GeV^2) \nonumber \\
 m_c      & = & 1.24 \pm 0.25 \pm 0.46~~GeV \nonumber \\
 \epsilon & = & 0.93 \pm 0.11 \pm 0.15 \nonumber
\eea
where the first error is statistical and the second error is 
systematic.  The contributions to the systematic error are shown
in Table~\ref{tab:tadams:cccharmerrs}.
The largest systematic error is from the (anti-)neutrino flux which is
expected to be reduced by further work.  However, the errors for
$\kappa$ and $\alpha$ are already statistics limited so the improvement
will be primarily for $m_c$ and $\epsilon$.  

\renewcommand{\arraystretch}{1.25}
\begin{table}
 \caption{Systematic errors for the parameters of the CC charm production
analysis. \label{tab:tadams:cccharmerrs}}
 \begin{center}
  \begin{tabular}{|lcccc|} \hline 
    & $\kappa$ & $\beta$ & $m_c$ & $\epsilon$ \\ \hline
  Hadron Calibration (1$\%$)              & 0.03 & 0.09 & 0.06 & 0.06 \\ \hline
  Muon Calibration (1$\%$)                & 0.02 & 0.14 & 0.11 & 0.05 \\ \hline
  Monte Carlo Calibration                 & 0.00 & 0.003 & 0.02 & 0.02 \\ \hline
  $R_L$ (20$\%$)                          & 0.00 & 0.10 & 0.03 & 0.01 \\ \hline
  $\pi/K$ decay 
  ($\nu$ 15$\%$ $\bar{\nu}$ 21$\%$)       & 0.02 & 0.17 & 0.01 & 0.08 \\ \hline
  Charm quark semi-muonic BF (10$\%$)     & 0.01 & 0.004 & 0.01 & 0.05 \\ \hline
  Monte Carlo Statistics                  & 0.02 & 0.11 & 0.05 & 0.02 \\ \hline
  Flux                                    & 0.03 & 0.28 & 0.44 & 0.08 \\ \hline \hline
  Total Systematic                        & 0.06 & 0.40 & 0.46 & 0.15 \\ \hline
  Statistics                              & 0.07 & 0.56 & 0.25 & 0.11 \\ \hline
  \end{tabular}
 \end{center}
\end{table}

\begin{figure}
 \centerline{\psfig{figure=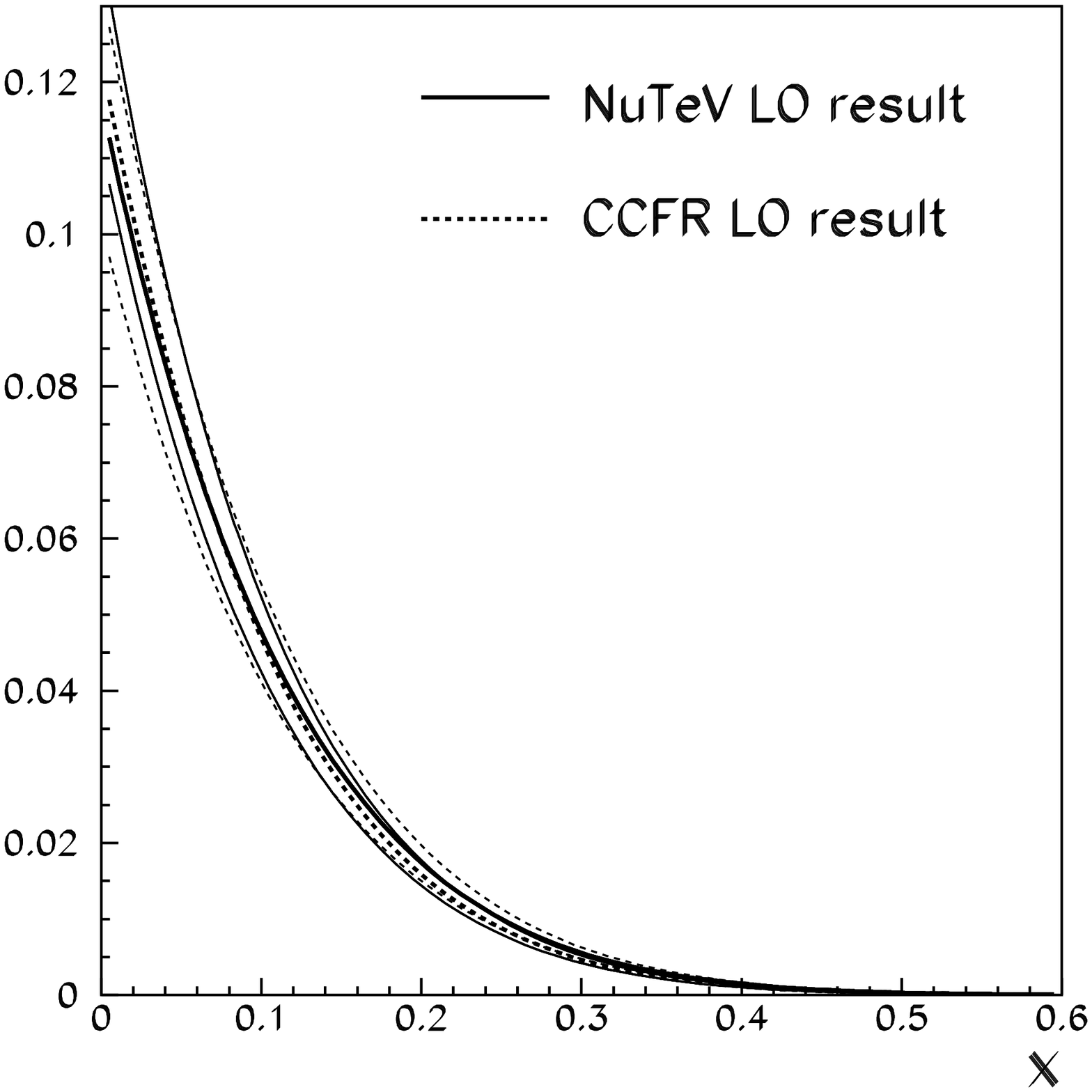,width=7.cm}}
 \caption{Comparison of the NuTeV (with error bands) result to the
CCFR result.
 \label{fig:tadams:ssea1}}
\end{figure}

\begin{figure}
 \centerline{\psfig{figure=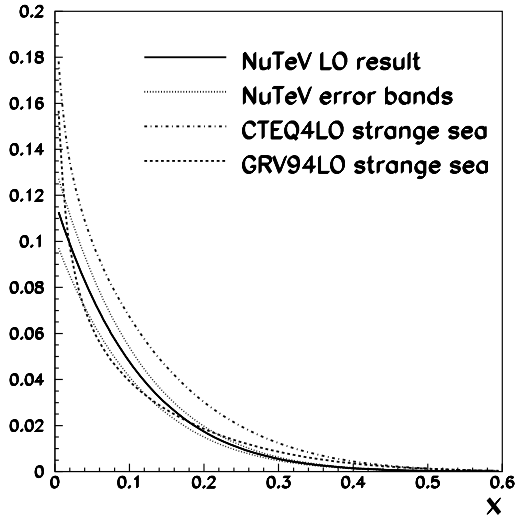,width=7.cm}}
 \caption{Comparison of the NuTeV (with error bands) result to
theoretical predictions from CTEQ and GRV.
 \label{fig:tadams:ssea2}}
\end{figure}

The results of this analysis for the strange sea compare favorably with 
previous measurements.  CHARM II,~\cite{bib:tadams:charmii}
CCFR~\cite{bib:tadams:rabin} and CDHS~\cite{bib:tadams:cdhs} have all
measured the size of the strange sea to be $\sim40\%$ of the non-strange
sea.  
The value of $\beta$ is consistent with CCFR's analysis which used
the same form for the momentum distribution.  
Figure~\ref{fig:tadams:ssea1} shows
a comparison of the results of the CCFR and NuTeV strange sea measurements.
There is excellent agreement between the two measurements.
Figure~\ref{fig:tadams:ssea2} compares the NuTeV result to theoretical
predictions from CTEQ and GRV.  The NuTeV result is lower than the
CTEQ prediction while in better agreement with GRV.


\section{Asymmetric Strange Sea \label{sec:tadams:asymssea}}

Predictions have been made that the nucleon strange sea
may have an asymmetry between the $s$ and $\bar{s}$ momentum
distributions.~\cite{bib:tadams:brodskyma}  If the nucleon has
a sizeable $s$ contribution at moderate $x$ it is possible for
the nucleon wavefunction to contain intrinsic $s$ states such
as $K^+ \Lambda$.  The Brodsky and Ma 
model~\cite{bib:tadams:brodskyma} predicts the $s$ momentum
distribution will be harder than the $\bar{s}$ distribution.

The CCFR collaboration has performed a test of this hypothesis.  The
$s$ and $\bar{s}$ momentum distributions were allowed to have
different shapes ($(1 - x)^\beta$ and $(1 - x)^{\beta^\prime})$)
while the total contributions were constrained to be equal:
\begin{eqnarray}
\int_{0}^{1} s(x) dx & = & \int_{0}^{1} \bar{s}(x) dx.
\end{eqnarray}

The fit to the CCFR data yielded a shape difference, 
$\Delta \beta = \beta - \beta^\prime$, of -0.46 $\pm$ 0.42 $\pm$ 0.76.
This indicates the $s$ and $\bar{s}$ momentum distribution are
consistent.  A 90$\%$ limit can be set on the difference:
\begin{eqnarray}
-1.9 < & \Delta \beta & < 1.0
\end{eqnarray}
The strange sea asymmetry is currently being explored by the NuTeV
collaboration.

\section{Neutral-Current Charm Production \label{sec:tadams:nccharm}}

Neutral-current (NC) charm production occurs via scattering off charm
quarks in the nucleon (Fig.~\ref{fig:tadams:feyn_diagrams}(b)).  Analysis
of this reaction probes the charm content of the nucleon.  While some
models allow for an intrinsic charm content of the nucleon,
this analysis only considers gluon splitting as its source for charm.

This analysis considers events which have a single muon with
the opposite charge from charged-current interactions of the selected
beam type (wrong-sign events).  This is possible because of NuTeV's 
low contamination of wrong type neutrinos in the beam and the
ability to select the beam type ($\nu$ or $\bar{\nu}$).

There are three sources of wrong-sign events other than NC charm
production.  The largest source is from charged-current interactions
initiated by wrong type neutrinos in the beam.
This contamination is less than $2\times10^{-3}$ of the
beam intensity.  Also included in this category are 
charged-current events where the sign of the muon is mis-identified
(generally due to a large scatter in the toroid spectrometer).  The
second largest source comes from two-muon events (see 
Section~\ref{sec:tadams:cccharm}) where the primary muon is not
measured because it is of low energy or exits the detector.  The
third source comes from NC events where a $\pi/K$ in the shower decays prior
to interacting.

\begin{figure}
 \centerline{\psfig{figure=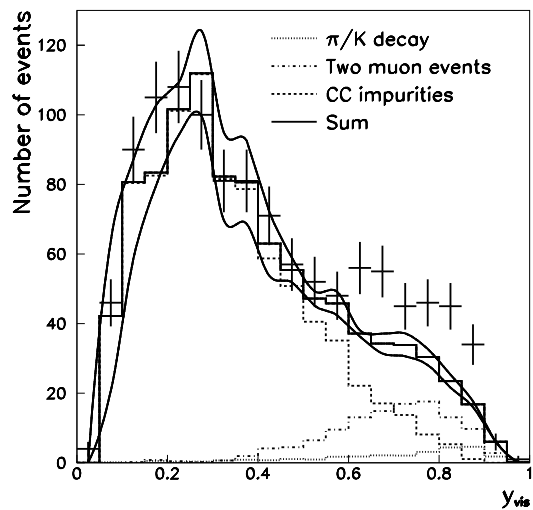,width=7.cm}}
 \caption{ NuTeV wrong-sign muon
data (points) compared to the sum of the background sources (solid histogram)
and systematic errors (curve).  The individual background components are
also shown as histograms.
 \label{fig:tadams:wsm1}}
\end{figure}

\begin{figure}
 \centerline{\psfig{figure=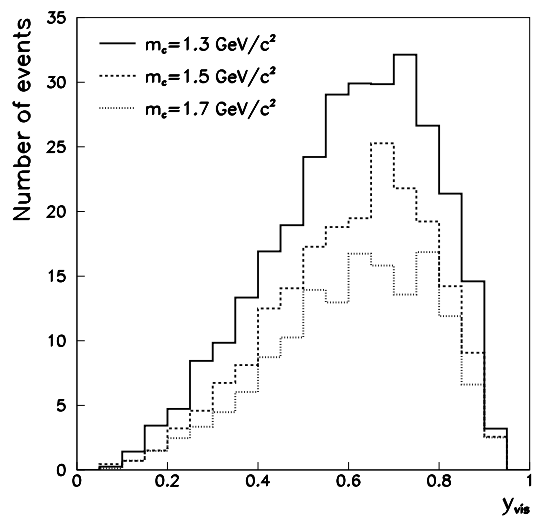,width=7.cm}}
 \caption{ Predictions of neutral-current charm
production for various values of $m_c$ = 1.3,1.5,1.7 ($\nu$ mode only).
 \label{fig:tadams:wsm2}}
\end{figure}

Figure~\ref{fig:tadams:wsm1} shows the wrong-sign muon data from
NuTeV (points) for the variable $y_{vis} = \frac{E_{HAD}}{E_{HAD} + E_\mu}$.  
The predictions for the individual background sources and
their sum are also shown.  A clear excess of events is seen at 
high $y_{vis}$.
Figure~\ref{fig:tadams:wsm2} shows predictions for neutral-current
charm production which peaks in the same region as the 
excess.~\cite{bib:tadams:nccharm}


\section{Summary}

The NuTeV sign-selected neutrino beams allow for improved studies of
the strange and charm contributions to the nucleon sea.  The strange sea
is observed to be in agreement with previous measurements with a size
which is $0.42 \pm 0.07 \pm 0.06$ the size of the average non-strange sea and
a shape $(1-x)^{8.5 \pm 0.56 \pm 0.39}$ (at $Q^2=16$ GeV$^2$).  
CCFR found no evidence for an asymmetric strange sea; NuTeV
is still investigating the subject.
The wrong-sign muon data shows a clear excess of events which is
consistent with neutral-current charm production.  

\section*{Acknowledgments}

We gratefully acknowledge the substantial contributions of the
staff of the Fermilab Beams and Particle Physics Divisions to the
construction and operation of the NuTeV beamlines and the Lab E
detector.  This work was sponsored in part by funding from the
U.S. Department of Energy and the National Science Foundation.

\end{document}